\documentclass{epl}

\usepackage[T1]{fontenc}
\usepackage{epsfig}

\title{Antiferromagnetism and single-particle properties in the
  two-dimensional half-filled Hubbard model: Slater {\it vs}
  Mott-Heisenberg }   
\shorttitle{Slater {\it vs} Mott-Heisenberg antiferromagnets}
\author{K. Borejsza and N. Dupuis}
\institute{ Laboratoire de Physique des Solides, CNRS UMR 8502, \\
  Universit\'e Paris-Sud, 91405 Orsay, France }
\date{December 17, 2002}

\pacs{71.10.Fd}{Lattice fermion models}
\pacs{71.10.Hf}{Non-Fermi-liquid ground states, electron phase diagrams
		and phase transitions in model systems}
\pacs{71.27.+a}{Strongly correlated electron systems, heavy fermions}

\begin{document}


\maketitle

\begin{abstract} 
We study antiferromagnetism and single-particle properties in the
two-dimensional half-filled Hubbard model at low
temperature. Collective spin fluctuations are governed by a non-linear
sigma model that we derive from the Hubbard model for any value of the
Coulomb repulsion. As the Coulomb repulsion increases, the ground
state progressively evolves from a Slater to a Mott-Heisenberg
antiferromagnet. At finite temperature, we find a metal-insulator
transition between a pseudogap phase at weak coupling and a
Mott-Hubbard insulator at strong coupling.
\end{abstract}


{\it Introduction.} 
The Hubbard model \cite{Gebhard97}
and its generalizations play a key role in the 
description of strongly correlated fermion systems such as high-$T_c$
superconductors, heavy fermions systems, or organic
conductors \cite{Imada98}. Despite its simplicity (the model is
defined by two parameters, the inter-site hopping amplitude $t$ and the
local Coulomb interaction $U$, and the symmetry of the lattice), exact
solutions or well-controlled approximations exist only in a few
special cases like in one-dimension (1D) or in the limit
of infinite dimension \cite{Gebhard97}.  

It is now well established that the ground state of the 2D half-filled
Hubbard
model on a square lattice has antiferromagnetic (AF) long-range
order. The origin of antiferromagnetism is believed to depend on the
strength of the Coulomb repulsion. At weak coupling ($U\ll t$), a
Fermi surface instability gives rise to an insulating spin-density-wave ground
state as first suggested by Slater \cite{Slater51}. In 2D, thermal
(classical) fluctuations preclude a finite-temperature transition
(Mermin-Wagner theorem) and the phase transition occurs at $T_{\rm
  N}=0$. Since the fermion spectral function is gapped at $T=0$, one 
expects, from a continuity argument, that it will exhibit a pseudogap
at finite temperature as a result of strong AF fluctuations
\cite{Vilk97}. At strong coupling ($U\gg t$), the system becomes a
Mott-Hubbard insulator below a temperature of the order of $U$. The
resulting local moments develop AF short-range order at a much
smaller temperature ($\sim J=4t^2/U$), and AF long-range order sets in
at $T_{\rm N}=0$.   

Although the Hartree-Fock (HF) theory gives a reasonable description of the
AF ground-state, it fails in 2D since it
predicts AF long-range order. Several alternative approaches, which do
satisfy the Mermin-Wagner theorem, have been proposed: Moriya's 
self-consistent-renormalized theory\cite{Moriya85}, the fluctuation
exchange approximation (FLEX) \cite{Bickers89}, and the two-particle
self-consistent approach \cite{Vilk97}. However, these approaches are
restricted to the weak to intermediate coupling regime
($U\sim 4t$). The strong-coupling regime
is usually understood from the Heisenberg model for which various methods are
available\cite{Auerbach94}.  

In this Letter, we describe a theoretical approach which provides a
unified view of antiferromagnetism and single-particle properties in
the 2D half-filled Hubbard model at low temperature 
(including $T=0$) and for any value of the Coulomb repulsion. It is
based on a non-linear sigma model (NL$\sigma$M) description of spin
fluctuations. Since it takes into account only directional
fluctuations of the AF order parameter, it is valid below a crossover
temperature $T_X$ which marks the onset of AF short-range order.

Besides its validity both at weak and strong coupling, our approach
differs from previous weak-coupling
theories\cite{Moriya85,Bickers89,Vilk97} by the fact that it is a 
low-temperature expansion ($0\leq T \ll T_X$). In particular, the
fermion spectral function is obtained from a spin-rotation-invariant
perturbative expansion around the 
(gapped) HF ordered state. This should be contrasted with
perturbative treatments applied to (gapless) free fermions interacting
with soft collective spin fluctuations where no small expansion parameter
is available \cite{Lee73}. In Ref. \cite{ND02}, one of the present authors
reported a calculation of the spectral function in the weak coupling limit
of the Hubbard model using a NL$\sigma$M description of spin fluctuations.
However, the limitations encountered by previous weak-coupling
theories could not be fully overcome.


{\it Non-linear sigma model.} 
The Hubbard model is defined by the Hamiltonian
\begin{equation}
H = -t \sum_{\langle {\bf r},{\bf r}'\rangle,\sigma} 
(c^\dagger_{{\bf r}\sigma} c_{{\bf r}'\sigma} + {\rm h.c.}) 
+U \sum_{\bf r} n_{{\bf r}\uparrow} n_{{\bf r}\downarrow} .
\label{Ham}
\end{equation}
$c^\dagger_{{\bf r}\sigma}$ ($c_{{\bf r}\sigma}$) 
creates (annihilates) a fermion of spin $\sigma$ at the lattice site
${\bf r}$. $  n_{{\bf r}\sigma}= c^\dagger_{{\bf r}\sigma}
c_{{\bf r}\sigma}$ and $\langle {\bf r},{\bf r}' \rangle$ denotes
nearest-neighbor sites. We take the lattice spacing equal to unity and
$\hbar=k_{\rm B}=1$. 

\begin{figure}
        \centerline{\psfig{file=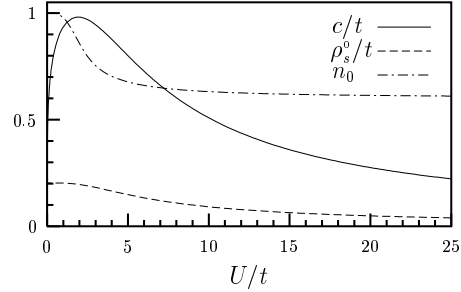,width=6cm,angle=0}}
        \caption{Spin-wave velocity $c$, bare spin stiffness
          $\rho_s^0$, and fraction $n_0$ of condensed bosons at $T=0$.}
\label{fig:rhocn0}
\end{figure}

We express charge and spin fluctuations in terms of auxiliary
fields. For this purpose, we write the interaction term in (\ref{Ham})
as $n_{{\bf r}\uparrow} n_{{\bf r}\downarrow}= [(  c^\dagger_{\bf r} 
c_{\bf r})^2 - (  c^\dagger_{\bf  r} \mbox{\boldmath$\sigma$} \cdot
{\bf\Omega}_{\bf r} c_{\bf r})^2]/4$ where ${\bf\Omega_r}$ is an
arbitrary unit vector \cite{Weng91}. $c_{\bf r}=(c_{{\bf r}\uparrow},
c_{{\bf r}\downarrow})^T$ and
$\boldsymbol{\sigma}=(\sigma_1,\sigma_2,\sigma_3)$ denotes the Pauli
matrices. Spin-rotation invariance is made
explicit by averaging the partition function over all directions of
${\bf\Omega_r}$. In a path-integral formalism, ${\bf\Omega_r}$ becomes
a time-dependent field. Decoupling the interaction term by means of
two real auxiliary fields, $\Delta_c$ and $\Delta_s$, the partition
function is then given by $Z=\int {\cal D}[c^\dagger,c]\int {\cal
D}[\Delta_c,\Delta_s,{\bf\Omega_r}]e^{-S}$ with the action
($\beta=1/T$) 
\begin{equation}
S = S_0 + \int _0^\beta d\tau  \sum_{\bf r} \left[  
\frac{\Delta_{c{\bf r}}^2+\Delta_{s{\bf r}}^2}{U} 
- c^\dagger_{\bf r}\bigl(i\Delta_{c{\bf r}}+\Delta_{s{\bf r}}
\mbox{\boldmath$\sigma$} \cdot {\bf\Omega}_{\bf r} \bigr) c_{\bf r} 
 \right] ,
\label{action1}
\end{equation}
where $c_{\bf r}$, $c^\dagger_{\bf r}$ are Grassmann variables.
$S_0$ is the action in the absence of interaction. Eq.~(\ref{action1})
defines an ``amplitude-direction'' representation, where the
AF order parameter field is given by $\Delta_{s{\bf
r}}{\bf\Omega_r}$. In the following, charge fluctuations ($\Delta_c$)
are considered at the saddle-point (i.e. HF) level:
$-i \Delta_{c \mathbf{r}} = (U/2) \left< c^{\dagger}_\mathbf{r}
c_\mathbf{r} \right> = U/2$.
Below the HF transition temperature $T_{\rm N}^{\rm HF}$, 
the amplitude of the order parameter takes a
well-defined value so that we can consider $\Delta_{s{\bf r}}$
within a saddle-point approximation, i.e. $\Delta_{s{\bf r}}=\Delta_0$
where the fluctuations of $\Delta_0$ are ignored. When $T\ll T_{\rm
N}^{\rm HF}$, $\Delta_0\sim
te^{-2\pi\sqrt{t/U}}$ for $U\ll t$ and tends to $U/2$ for $U\gg
t$. Below the crossover temperature $T_X$ (to be defined more
precisely later) which
marks the onset of AF short-range order, the ${\bf\Omega_r}$ field
can be parametrized by ${\bf\Omega}_{\bf r}=(-1)^{\bf r} {\bf n}_{\bf
  r}(1-{\bf L}^2_{\bf r})^{1/2}+{\bf L}_{\bf r}$, where the
N\'eel field ${\bf n_r}$  ($|{\bf n_r}|=1$) is assumed to be slowly
varying \cite{Haldane83}. The 
small canting vector ${\bf L_r}$, orthogonal to ${\bf n_r}$, takes
account of local ferromagnetic fluctuations. It turns out to be convenient to
introduce a pseudo-fermion $\phi_{{\bf r}\sigma}$ whose spin is
quantized along the (fluctuating) N\'eel field. $\phi_{\bf
  r}=(\phi_{{\bf r}\uparrow},\phi_{{\bf r}\downarrow})^T$ is
defined by $c_{\bf r}=R_{\bf r}\phi_{\bf r}$ where $R_{\bf r}$ is a
site- and time-dependent SU(2)/U(1) matrix satisfying $R_{\bf r}\sigma_3
R^\dagger_{\bf r}=\boldsymbol{\sigma}\cdot {\bf n_r}$. The action can
then be written as  
\begin{eqnarray}
                S = 
                \int_0^{\beta}d\tau \sum_{\mathbf{r}}
                \phi_{\mathbf{r} }^{\dagger} 
                \Big\lbrace
                        \partial_{\tau} -  A_{0 \mathbf{r}}
                -2 t \sum_{\mu = x,y}
                        \cos( -i \partial_{\mu} - A_{\mu \mathbf{r}} ) 
	- \Delta_0
                        [
                                (-1)^{\mathbf{r}} 
                                \sigma_3
                                \sqrt{ 1 - \mathbf{l}_{\mathbf{r} }^2 }
                                        + 
                                \mathbf{l}_{\mathbf{r} } 
                                \cdot
                                \mbox{\boldmath$\sigma$} ] 
		\Big\rbrace 
                        \phi_{ \mathbf{r} } ,
\label{action2}
\end{eqnarray}
where we have introduced the SU(2) gauge field $A_{0{\bf
    r}}=-R^\dagger_{\bf r}\partial_\tau R_{\bf r}$, $A_{\mu{\bf
    r}}=iR^\dagger_{\bf r}\partial_\mu R_{\bf r}$ ($\mu=x,y$), and the
rotated canting field ${\bf l_r}={\cal R}^{-1}_{\bf r}{\bf
L_r}$. Here ${\cal R}_{\bf r}$ is the SO(3) element associated to
$R_{\bf r}$ which maps $\hat {\bf z}$ onto ${\bf n_r}$. In
Eq.~(\ref{action2}), both ${\bf l}$ and $A_\mu$ are small, since the
gauge field is of order $\partial_\mu {\bf n}$. The effective action of
the N\'eel field is obtained by expanding (\ref{action2}) to second
order in these variables and integrating out the fermions
and the canting field $\mathbf{l}_\mathbf{r}$. Skipping
technical details, we obtain
\begin{equation}
                S_{ \mathrm{NL}\sigma\mathrm{M} }[\mathbf{n}] 
        =
                \frac{\rho_s^0}{2} 
                \int_0^{\beta} d \tau \int d^2 r 
                \left[
                        \frac{(\partial_\tau\mathbf{n}_{\bf r})^2}{c^2}
                        + 
                        ( \boldsymbol{\nabla} \mathbf{n}_{\bf r} )^2
                \right] ,
\label{nlsm}
\end{equation}
where we have taken the continuum limit in real space. 
The bare spin stiffness $\rho_s^0$ and the spin-wave velocity $c$ are given by 
$\rho_s^0 = \epsilon_c/8$ and $
        c^2 = (\epsilon_c/2)(\chi^{-1}_{\perp} -U/2)$,
where $\epsilon_c$ is the absolute value of the (negative) kinetic
energy per site and $\chi_\perp$ the transverse spin susceptibility in
the HF ground state (Fig.~\ref{fig:rhocn0}). In the weak-coupling
limit, AF short-range order 
cannot be defined at length scales smaller than $\xi_0\sim t/\Delta_0$
which corresponds to the size of bound particle-hole pairs in the HF
ground state. Thus Eq.~(\ref{nlsm}) should be supplement with a cutoff
$\Lambda \sim {\rm min}(1,\xi_0^{-1})$ in momentum space. The
NL$\sigma$M (\ref{nlsm}) was first obtained
by Schulz \cite{Schulz95}. In the limit $U\gg t$, it reproduces the
result obtained from the Heisenberg model with an exchange coupling
$J=4t^2/U$ \cite{Auerbach94}. 

\begin{figure}
         \centerline{\psfig{file=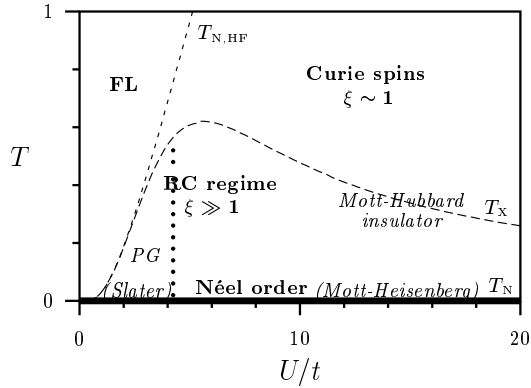,width=7.cm,angle=0}}
        \caption{Phase diagram of the 2D half-filled Hubbard model. 
                All lines, except $T_{\rm N}=0$ (thick solid line),
                are crossover lines. The 
                NL$\sigma$M description is valid below $T_X$. 
		FL - Fermi liquid phase, PG - pseudogap phase. The
                vertical dotted line indicates the finite-temperature
                metal-insulator transition obtained from the criterion
                $\rho(\omega=0)=0$. }
\label{fig0diag0phases0magentique}
\end{figure}

{\it Magnetic phase diagram.}
We solve the NL$\sigma$M within the
CP$^1$ representation, where the N\'eel field is
expressed in terms of two Schwinger bosons: ${\bf n_r}=z^\dagger_{\bf
  r} \boldsymbol{\sigma} z_{\bf r}$ ($z_{\bf r}=(z_{{\bf r}\uparrow},
z_{{\bf r}\downarrow})^T$) with $z^\dagger_{\bf r}z_{\bf r}=1$. When the
CP$^1$ representation is generalized to the CP$^{N-1}$ representation
by introducing $N$ bosons $z_{{\bf r}\sigma}$ ($\sigma=1,...,N$), the
NL$\sigma$M can be solved exactly by a saddle-point approximation
in the $N \rightarrow \infty$ limit \cite{Auerbach94}. At zero temperature,
there is a quantum critical point at $g=g_c=4\pi/\Lambda$ 
between a phase with AF long-range order and a (quantum) disordered
phase. $g=c/\rho_s^0$ is the coupling constant of the
NL$\sigma$M. In the ordered phase ($g<g_c$), a fraction
$n_0=1-g/g_c$ ($0\leq n_0\leq 1$) of the bosons condenses in the mode ${\bf
q}=0$. $n_0$ determines the mean-value of the N\'eel field: $|\langle
{\bf n_r}\rangle |=n_0$. We find that there is
AF long-range order for any value of the Coulomb repulsion in the
ground state of the 2D half-filled Hubbard 
model. At weak coupling, $1-n_0=g/g_c\sim e^{-2\pi\sqrt{t/U}}$ is
exponentially small.  By an appropriate choice of the cutoff
$\Lambda$, we reproduce the result $n_0\simeq 0.6$ for $U\gg t$ as
obtained from the 2D Heisenberg model on a square lattice
\cite{Manousakis91} (Fig.~\ref{fig:rhocn0}). At finite temperature, the AF
long-range order is suppressed ($n_0=0$). However, the AF correlation
length remains exponentially 
large: $\xi\sim (c/T)e^{2\pi\rho_s/T}$, where
$\rho_s=\rho_s^0(1-g/g_c)$ is the zero temperature spin stiffness. This regime, which
is dominated by classical (thermal) fluctuations (since $c/\xi\ll T$),
is known as the renormalized classical regime. The NL$\sigma$M
description is valid for $T < T_X$ when amplitude fluctuations of
the AF order parameter are frozen and the assumption of AF short-range
order holds (i.e. $\xi\gg \Lambda^{-1}$).  At weak
coupling $T_X\sim T_{\rm N}^{\rm HF}$, while $T_X\sim J$ at
strong coupling. The phase diagram is shown in
Fig.~\ref{fig0diag0phases0magentique}. Above
$T_{\mathrm{N}}^{\mathrm{HF}}$, spin fluctuations are not important
and we expect a Fermi liquid (FL) behavior. Between
$T_{\mathrm{N}}^{\mathrm{HF}}$ and $T_\mathrm{X}$ (a regime which
exists only in the strong-coupling limit), local moments form
($\xi_0\sim 1$) but with no AF short-range order (Curie spins: $\xi
\sim 1$). Below $T_\mathrm{X}$, the system 
enters a renormalized classical regime of spin fluctuations where the AF
correlation length becomes exponentially large. AF
long-range order sets in at $T_\mathrm{N} = 0$. Although there is a smooth
evolution of the magnetic properties as a function of $U$, the physics is quite
different for $U \ll t$ and $U \gg t$. This is shown below by
discussing the fermion spectral function. The main 
conclusions are summarized in Fig.~\ref{fig0diag0phases0magentique}. 

{\it Fermion spectral function.} 
Now we consider the effect of long-wavelength spin fluctuations on the
fermion spectral  function. The fermion Green's function
${\cal G}({\bf r},\tau;{\bf r}',\tau')=-\langle c_{\bf r}(\tau)
c^\dagger_{{\bf r}'}(\tau')\rangle$, written here as a $2\times 2$
matrix in spin space, is computed using $c_{\bf r}=R_{\bf r}\phi_{\bf
r}$ with $(R_{\bf r})_{\uparrow\uparrow}=(R_{\bf
r})^*_{\downarrow\downarrow}=z_{{\bf r}\uparrow}$ and 
$(R_{\bf r})_{\downarrow\uparrow}=-(R_{\bf
r})^*_{\uparrow\downarrow}=z_{{\bf r}\downarrow}$. Integrating first
the pseudo-fermions and the canting field $\mathbf{L}_\mathbf{r}$,
we can write the Green's function as
\begin{equation}
                \mathcal{G}(1,2) = 
                \frac{1}{Z}
                \int \mathcal{D}[z]
                e^{ -S_{\mathrm{NL}\sigma\mathrm{M}} [z ] }
                R_1
                \mathcal{G}(1,2|z)
                R_2^{\dagger} ,
\label{gf0}
\end{equation}
where we use the shorthand notation $i\equiv ({\bf
r}_i,\tau_i)$. $\mathcal{G}(1,2|z)$ is the pseudo-fermion
propagator calculated for a given configuration of the bosonic field
$z$. The action (\ref{action2}), when expanded in powers of $A_\mu$
and ${\bf L_r}$, can be written as $S_{\rm HF}[\phi]+S'[z,\phi,{\bf
L}]$. It describes HF pseudo-fermions interacting with spin
fluctuations {\it via} the action $S'$. Since the HF pseudo-fermions
are gapped, we expect a perturbative expansion in $S'$ to be
well-behaved. To leading order
$\mathcal{G}(1,2|z)=\mathcal{G}^{\mathrm{HF}}(1,2)$ with ${\cal
  G}^{\rm HF}$ the HF Green's function. From (\ref{gf0}),
we then obtain 
\begin{eqnarray}
\mathcal{G}_{\sigma} ( \mathbf{k} , \mathbf{k}' , \omega ) 
        &=&        - \frac        { 
                        2 \delta_{ \mathbf{k} , \mathbf{k}' }
                        }
                        {
                        \beta
                        }
                \sum_{\omega_{\nu}}
                \int_{ \mathbf{q} }
                \mathcal{G}_{\sigma}^{\mathrm{HF}}
                  ( \mathbf{k-q} , \mathbf{k-q} , \omega - \omega_{\nu} ) 
                 \overline{\mathcal{D}}( \mathbf{q} , \omega_{\nu} )
        +
                n_0 
                \mathcal{G}_{\sigma}^{\mathrm{HF}}( \mathbf{k} ,
        \mathbf{k}' ,  \omega )  , \label{gf} \\ 
	\mathcal{G}_{\sigma}^{\mathrm{HF}}(\mathbf{k} , \mathbf{k}' , \omega )
	&=&
		-
		\delta_{ \mathbf{k} , \mathbf{k}' } 
		\frac	{ i \omega + \epsilon_\mathbf{k} }
			{ \omega^2 + E_\mathbf{k}^2 }
		+
		\delta_{ \mathbf{k} , \mathbf{k}'+ \boldsymbol{\pi} } 
		\frac	{ \sigma  \Delta }
			{ \omega^2 + E_\mathbf{k}^2 } , \quad
                 \bar {\cal D} ({\bf q},\omega_\nu) =
        -\frac{gc/2}{\omega_\nu^2+\omega^2_{\bf q}} , 
\end{eqnarray}
where $\int_{\bf q}\equiv \int_{-\pi}^{\pi}\frac{dq_x}{2\pi}
\int_{-\pi}^{\pi}\frac{dq_y}{2\pi}$, $\boldsymbol{\pi}=(\pi,\pi)$,
and $\omega$ ($\omega_\nu$) 
denotes a fermionic (bosonic) Matsubara frequency.
$\bar {\cal D}({\bf q},\omega_\nu)$ is the Schwinger boson propagator (for
${\bf q}\neq 0$) obtained from the saddle-point solution of the
NL$\sigma$M. 
Here $\omega_{\bf q}=c({\bf q}^2+\xi^{-2}/4)^{1/2}$ and
$E_{\bf k}=(\epsilon^2_{\bf k}+\Delta_0^2)^{1/2}$, with $\epsilon_{\bf
  k}=-2t(\cos k_x+\cos k_y)$ the energy of the free fermions. 
At finite temperature, $n_0$ vanishes so that the fermion Green's
function is spin-rotation and translation invariant. From (\ref{gf}),
we obtain the spectral function $\mathcal{A} ( \mathbf{k} , \omega
)=-\pi^{-1}{\rm Im}{\cal G}_\sigma({\bf k},{\bf k},i\omega\to \omega+i0^+)$: 
\begin{eqnarray}
                \mathcal{A} ( \mathbf{k} , \omega )
        &=&
                \mathcal{A}_{\mathrm{inc}} ( \mathbf{k} , \omega )
                + n_0 \mathcal{A}_{\mathrm{HF}} ( \mathbf{k} , \omega )
        \label{eq0fc0spec0decomp} , \\
                \mathcal{A}_{\mathrm{inc}} ( \mathbf{k} , \omega )
        &=&
                \int_{\mathbf{q}} \frac{g c}{2 \omega_q}
                \Bigl\lbrace \left[
                        n_B( \omega_\mathbf{q} ) + n_F( -E_{\mathbf{k-q}} )
                \right]
                \left[
                        u_{\mathbf{k-q}}^2 
                        \delta( \omega - \omega_{\mathbf{q}} - E_{\mathbf{k-q}} )
                        +
                        v_{\mathbf{k-q}}^2 
                        \delta( \omega + \omega_{\mathbf{q}} + E_{\mathbf{k-q}} )
                \right]
        \nonumber \\
        && +
                \left[
                        n_B( \omega_\mathbf{q} ) + n_F( E_{\mathbf{k-q}} )
                \right]
                \left[
                        u_{\mathbf{k-q}}^2 
                        \delta( \omega + \omega_{\mathbf{q}} - E_{\mathbf{k-q}} )
                        +
                        v_{\mathbf{k-q}}^2 
                        \delta( \omega - \omega_{\mathbf{q}} + E_{\mathbf{k-q}} )
                \right]\Bigr\rbrace ,
        \label{eq0fc0spec0inc} 
\end{eqnarray}
where $n_F(\omega)$ and $n_B(\omega)$ are the usual Fermi and Bose occupation 
numbers and
$\mathcal{A}_{\mathrm{HF}}$ the HF spectral function:
$
                \mathcal{A}_{ \mathrm{HF} } ( \mathbf{k} , \omega )     
        =
                u_{\mathbf{k}}^2
                \delta( \omega - E_{\mathbf{k}} )
                +
                v_{\mathbf{k}}^2
                \delta( \omega + E_{\mathbf{k}} )
        \label{eq0fc0spec0hf}  
$,
where
$                u_{\mathbf{k}}^2,v_{\mathbf{k}}^2
        =
                \frac{1}{2}
                \left(
                        1 \pm \frac{\epsilon_\mathbf{k}}{E_\mathbf{k}}
                \right)
$. 
The normalization of the spectral function, $\int d\omega
\mathcal{A}(\mathbf{k},\omega)=1$, follows from the saddle-point
equation of the NL$\sigma$M. 

\begin{figure}
	\centerline{\psfig{file=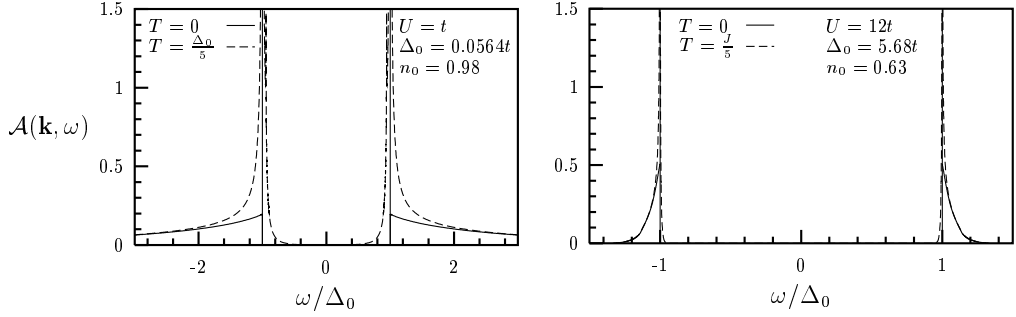,width=13.5cm,angle=0}}
 	\caption{Left: spectral function $\mathcal{A}(
        	\mathbf{k}, \omega)$  in the weak-coupling limit $U = t$ for $T=0$
        	(Slater antiferromagnet)  and $T=\Delta_0/5$ (pseudogap phase). 
        	Right:	spectral function in the strong-coupling regime $U = 12 t$ 
		for $T=0$ (Mott-Heisenberg antiferromagnet) and $T=J/5$ 
		(Mott-Hubbard insulator).
		$\mathbf{k} = (\pi/2,\pi/2)$. For $T=0$, the vertical lines
        	represent Dirac peaks of weight $n_0/2$. Note the difference 
		in the energy scale, which is fixed by $\Delta_0$, between 
		the two figures.
        	}
        \label{fig0fc0spec0tnulle0ufaible}
\end{figure}

\medskip
        
The spectral function is shown in Fig.~\ref{fig0fc0spec0tnulle0ufaible}.
At weak coupling and zero temperature, our theory describes
a Slater   
antiferromagnet. The AF gap $2\Delta_0\sim t e^{-2\pi\sqrt{t/U}}$ is
exponentially small. There are well-defined Bogoliubov quasiparticles (QP's)
with energy $\pm E_{\bf k}$, as in the
HF theory, but their spectral weight is reduced by a factor $n_0$ due
to quantum spin fluctuations. The remaining weight ($1-n_0\sim
e^{-2\pi\sqrt{t/U}}\ll 1$) is carried by an incoherent excitation
background at higher energy ($|\omega|>E_{\bf k}$). 
As $U$ increases, the Slater antiferromagnet progressively
evolves into a Mott-Heisenberg antiferromagnet with a large AF gap
and a significant fraction of spectral weight transferred from the
Bogoliubov QP's to the incoherent excitation background. At strong
coupling, the AF gap $2\Delta_0\simeq U$ and $n_0\simeq 0.6$. 

At finite temperature, ${\cal A}({\bf k},\omega)$ exhibits two
broadened peaks at the HF QP energy $\pm E_{\bf k}$.
In the vicinity of the peak at $E_{\bf k}$,
the spectral function is well approximated by
$
{\cal A}^>_{\rm peak}({\bf k},\omega) = u^2_{\bf k} \frac{g}{4\pi c}
n_B(|\omega-E_{\bf k}|) .
$
The peak has a width of the order of the temperature and therefore
corresponds to 
incoherent excitations. We find that the spectral weight of the peak,
$\int d\omega {\cal A}^>_{\rm peak}({\bf k},\omega)$, is
temperature independent and equal to $u^2_{\bf k}n_0$, which is
nothing else but the Bogoliubov QP weight in the ground state. We
conclude that the peak at $E_{\bf k}$ is an incoherent precursor
of the zero-temperature Bogoliubov QP peak. As the temperature
decreases, it retains its spectral weight but becomes sharper and
sharper, and eventually becomes a Dirac peak at $T=0$. As expected,
the weak coupling pseudogap continuously evolves into the AF gap
when $T \rightarrow 0$.

Eq.~(\ref{eq0fc0spec0inc}) shows that the contribution to ${\cal A}({\bf
  k},\omega)$ at low energy involves the Bose
occupation number $n_B(\omega_{\bf q})$. This indicates that the
low-energy fermion states ($0\leq|\omega|<E_{\bf k}$) are due to thermal
bosons, i.e. thermally excited spin fluctuations. 
A fermion added to the system with momentum 
$\mathbf{k}$ and energy $|\omega| < E_\mathbf{k}$ can propagate by absorbing a thermal 
boson of energy $\omega_\mathbf{q}$ and emitting a pseudo-fermion with energy 
$E_{\mathbf{k}-\mathbf{q}} = \omega + \omega_\mathbf{q}$. 
The lowest fermion energies are obtained by solving 
$\omega=E_{\mathbf{k}-\mathbf{q}}-\omega_\mathbf{q}$ 
(or  $\omega=-E_{\mathbf{k}+\mathbf{q}}+\omega_\mathbf{q}$). 
In the weak coupling limit (Fig.~\ref{fig0fc0spec0tnulle0ufaible}), 
$\mathrm{max}_\mathbf{q} (\omega_\mathbf{q}) = c \Lambda \sim 2 \Delta_0$ 
and 
$E_{\mathbf{k}-\mathbf{q}} \sim E_\mathbf{k}$. 
Thus there is spectral weight at zero energy: the spectral function and the 
density of states exhibit a pseudogap.
Nevertheless, the density of states
$\rho(\omega)=\int_{\bf k} {\cal A}({\bf k},\omega)$ remains
exponentially small at low energy:  
$\rho(\omega) \sim e^{-\Delta_0/T} \cosh(
\omega/T ), \,\, |\omega| \ll \Delta_0$. This result differs from
pseudogap theories based on Gaussian spin fluctuations which find a
much weaker suppression of the density of states \cite{Lee73}. It
bears some similarities with the result obtained by Bartosch and
Kopietz for fermions coupled to classical phase fluctuations in
incommensurate Peierls chains \cite{Bartosch00}. 
In the low temperature regime dominated by directional fluctuations
of the order parameter, the suppression of the density of states
at low energy is indeed expected to be exponential.
In the strong coupling limit, since 
$E_\mathbf{k}\sim U/2$ and  $c\Lambda \sim J\ll U/2$, there is a gap (of order $U/2$) 
in the spectral function and the density of states.
Thermally excited spin fluctuations reduce the
zero-temperature AF gap $U/2$ by a small amount of the order of $J$. The system 
is a Mott-Hubbard insulator
(Fig.~\ref{fig0diag0phases0magentique}).  

\begin{figure}
        \centerline{\psfig{file=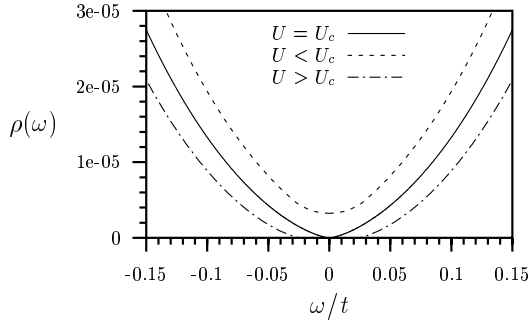,width=7.cm,angle=0}}
        \caption{Density of states $\rho(\omega)$ at low energy.
 	As the Coulomb repulsion $U$ increases through a critical 
	value $U_c \simeq 4.25 t$ the pseudogap becomes a Mott-Hubbard
	gap [see the vertical dotted line in Fig.~\ref{fig0diag0phases0magentique}].
        }
\label{fig0mit0dos}
\end{figure}

We therefore conclude that our approach predicts a finite-temperature
metal-insulator transition between a pseudogap phase and a Mott-Hubbard insulator as
the strength of the Coulomb interaction increases
(Fig.~\ref{fig0diag0phases0magentique}): at a critical value $U_c$,
the density of states at zero energy $\rho(\omega=0)$ 
vanishes and the pseudogap becomes a Mott-Hubbard gap (Fig.~\ref{fig0mit0dos}).
$U_c$ is obtained by equating
the minimum energy $\Delta_0$ of a HF fermion to the maximum energy
of a Schwinger boson $\sqrt{ m^2 + c^2 \Lambda^2}$. For $T \rightarrow 0$
the result is $U_c \simeq 4.25 t$.
However, being a low-energy theory,
the NL$\sigma$M does not allow us to describe accurately the high-energy Schwinger
bosons (with  $|\mathbf{q} | \sim \Lambda$) and in turn the low-energy
fermion excitations. In particular, the critical value of $U$ calculated 
above and the precise form of the density of states near zero energy, plotted
in Fig.~\ref{fig0mit0dos}, depend on the cut-off procedure used in the
NL$\sigma$M. Note also that we do
not know at which temperature and how the metal-insulator transition ends.


{\it Conclusion.} 
We have presented a low-temperature approach to the 2D half-filled
Hubbard model 
which allows us to study both collective spin fluctuations and single-particle
properties for any value of the Coulomb repulsion $U$. At zero
temperature, it describes the evolution from a Slater to a
Mott-Heisenberg antiferromagnet. At finite temperature, it predicts a
metal-insulator transition between a pseudogap phase at weak coupling
and a Mott-Hubbard insulator at strong coupling. Since the charge
auxiliary field $\Delta_c$ is considered at the HF level, some aspects of the
Mott-Hubbard localization are not taken into account. In particular,
at intermediate coupling $U\sim 8t$, we expect both Bogoliubov QP bands
(or precursors thereof at finite temperature) and Mott-Hubbard bands
in the spectral function \cite{Vilk97}. The Mott-Hubbard bands have a
purely local origin, independent of the Fermi surface geometry, and
should show up at $\omega\sim \pm U/2$ (with $U/2>\Delta_0$) in the spectral
function. Nevertheless, we believe that our theory captures the main
features of the physics of the 2D half-filled Hubbard model.

On the basis of a numerical calculation (dynamical cluster
approximation), Moukouri and Jarrell have called into question the
existence of a Slater mechanism in the 2D Hubbard model\cite{Moukouri01}. 
Using the criterion $\rho(\omega=0)<10^{-2}/(2t)$ to identify the Mott
insulating 
phase, they concluded that the system is always insulating at low (but finite)
temperature even in the weak-coupling limit. From the same criterion
($\rho(\omega=0)<10^{-2}/(2t)$), we obtain a similar line in the ($U,T$)
plane as Moukouri and Jarrell. This shows that the 
numerical results of Ref.~\cite{Moukouri01} are not in contradiction
with the existence of a Slater scenario at weak coupling, but reflect
the exponential suppression of the density of states due to the
presence of a pseudogap.


\vspace{-0.03cm}
\acknowledgments
\vspace{-0.3cm}
We thank A.-M.S. Tremblay for useful correspondence and S. Pairault for a critical
reading of the manuscript.


\end{document}